\documentclass[aps,prl,twocolumn,showpacs,floatfix,superscriptaddress,preprintnumbers]{revtex4}
\usepackage{amsmath}
\usepackage{amssymb}
\usepackage{epsfig}
\usepackage{graphicx}
\usepackage{stmaryrd}
\usepackage{pstricks}
\usepackage{pst-coil}
\DeclareMathAlphabet{\mathsc}{OT1}{cmr}{m}{sc}

\renewcommand{\baselinestretch}{1.55}

\def\10{$SO(10)$}
\def\21{SU(2) $\otimes$ U(1) }

\def\422{$SU(4) \otimes SU(2) \otimes SU(2)$}
\def\321{SU(3) $\otimes$ SU(2) $\otimes$ U(1)}

\def\lsim{\raise0.3ex\hbox{$\;<$\kern-0.75em\raise-1.1ex\hbox{$\sim\;$}}}
\def\gsim{\raise0.3ex\hbox{$\;>$\kern-0.75em\raise-1.1ex\hbox{$\sim\;$}}}
\def\vev#1{\left\langle #1\right\rangle}

\newcommand{\AddrAHEP}{%
  AHEP Group, Institut de F\'{\i}sica Corpuscular --
  C.S.I.C./Universitat de Val{\`e}ncia \\
  Edificio Institutos de Paterna, Apt 22085, E--46071 Valencia, Spain}

\renewcommand{\baselinestretch}{1.14}
 \newcommand{\ba}{\begin{array}}
\newcommand{\ea}{\end{array}}
\relax
\def\321{$SU(3)\times SU(2)\times U(1)$}
\def\mnu{{\cal M}_{\nu }}
\newcommand{\Sol}  {\textrm{sol}}

\newcommand{\Atm}  {\textrm{atm}}

\newcommand{\Dms}  {\Delta m^2_\Sol}
\newcommand{\Dma}  {\Delta m^2_\Atm}

\begin{document}
\preprint{IFIC/08-19}
\renewcommand{\Huge}{\Large}
\renewcommand{\LARGE}{\Large}
\renewcommand{\Large}{\large}
\def \znbb {$0\nu\beta\beta$ }
\def \nbb {$\beta\beta_{0\nu}$ }
\title{Tri-bimaximal neutrino mixing and neutrinoless double beta
  decay}
\author{M.~Hirsch} \email{mahirsch@ific.uv.es} 
\author{S.~Morisi} \email{ Stefano.Morisi@ific.uv.es}
\author{J.~W.~F.~Valle} \email{valle@ific.uv.es}
\affiliation{\AddrAHEP}

\date{\today}

\begin{abstract}

  We present a tri-bimaximal lepton mixing scheme where the
  neutrinoless double beta decay rate has a lower bound which
  correlates with the ratio $\alpha \equiv \Dms/\Dma$ well determined
  by current data, as well as with the unknown Majorana CP phase
  $\phi_{12}$ characterizing the solar neutrino sub-system. For the
  special value $\phi_{12}=\frac{\pi}{2}$ (opposite CP-sign neutrinos)
  the \nbb rate vanishes at tree level when $\Dms/\Dma = 3/80$, only
  allowed at $3\sigma$. For all other cases the rate is nonzero, and
  lies within current and projected experimental sensitivities
  close to $\phi_{12}=0$. We
  suggest two model realizations of this scheme in terms of an $A_4
  \times Z_2$ and $A_4 \times Z_4$ flavour symmetries.

\end{abstract}


\maketitle


Current neutrino oscillation
data~\cite{fukuda:2002pe,ahmad:2002jz,araki:2004mb,Kajita:2004ga,ahn:2002up,Collaboration:2007zza,kamland:07}
indicate a peculiar pattern~\cite{Maltoni:2004ei} of neutrino masses
and mixings quite at variance with the structure of the
Cabibbo-Kobayashi-Maskawa quark mixing matrix~\cite{Kobayashi:1973fv}.
However they do not yet fully determine the absolute scale of neutrino
masses nor shed any light on the issue of leptonic CP violation, two
demanding challenges left for future experiments.

Lacking a basic theory for the origin of mass one needs theoretical
models restricting the pattern of fermion masses and mixings and
providing guidance for future experimental searches.
An attractive phenomenological ansatz for leptons is the
Harrison-Perkins-Scott (HPS) mixing~\cite{Harrison:2002er}
\begin{equation}
\label{eq:HPS}
U_{\textrm{HPS}} = 
\begin{pmatrix}
\sqrt{2/3} & 1/\sqrt{3} & 0\\
-1/\sqrt{6} & 1/\sqrt{3} & -1/\sqrt{2}\\
-1/\sqrt{6} & 1/\sqrt{3} & 1/\sqrt{2}
\end{pmatrix}
\end{equation}
which predicts the following values for the lepton mixing angles:
$\tan^2\theta_{\Atm}=1$, $\sin^2\theta_{\textrm{Chooz}}=0$ and
$\tan^2\theta_{\Sol}=0.5$, providing a good first approximation to the
values~\cite{Maltoni:2004ei} indicated by neutrino oscillation
experiments~\cite{fukuda:2002pe,ahmad:2002jz,araki:2004mb,Kajita:2004ga,ahn:2002up}.

As noted earlier~\cite{Ma:2004zv}, when the charged lepton mass matrix $M_l$ obeys
$$
M^lM^{l\dagger}=U_\omega\,M^{l2}_{diag}\,U_\omega^\dagger;
$$
where $U_\omega$ is the ``magic'' unitary matrix
$$U_\omega=
\frac{1}{\sqrt3}\left(\begin{array}{ccc}
1&1&1\\
1&\omega&\omega^2\\
1&\omega^2&\omega\\
\end{array}\right),
$$
and the neutrino mass matrix has the form
$$
M_\nu\sim
\left(\begin{array}{ccc}
A&0&0\\
0&B&C\\
0&C&B
\end{array}\right),
$$
the resulting lepton mixing matrix has exactly the tri-bimaximal
structure given in Eq.~(\ref{eq:HPS}).

Here we consider schemes where neutrinos get mass {\sl a la seesaw},
defined by the following mass matrices,
$$
M^l\sim
\left(\begin{array}{ccc}
\alpha &\beta &\gamma\\
\gamma & \alpha & \beta\\
\beta &\gamma  &\alpha\\
\end{array}\right)=U_\omega\,M^l_{diag}\,U_\omega^\dagger;
$$
$$m_D\sim
\left(\begin{array}{ccc}
a &0&0\\
0& a & b\\
0& b & a\\
\end{array}\right);
M_R\sim
\left(\begin{array}{ccc}
1&0&0\\
0&1&0\\
0&0&1\\
\end{array}\right)
$$
This ``texture'' constitutes a new ansatz for the lepton sector that
can be realized (see below) in the framework of $A_4$-based flavour
symmetry models.
The assumed symmetry of the Dirac mass term holds in $SO(10)$ models
where it comes from a 16\,16\,10 Yukawa coupling.
In contrast with other existing tri-bimaximal $A_4$ based schemes, the
gauge singlet seesaw mass term characterizing the heavy right-handed
neutrinos is also a flavour singlet, instead of the neutrino Dirac
mass term.
This makes the scheme extremely predictice, as it involves as free parameters only the
two modulii and the relative phase between $a$ and $b$.

After the seesaw mechanism, one obtains the effective light neutrino
mass matrix $M_\nu$ given as
\begin{equation}\label{mnu1}
M_\nu=m_D\frac{1}{M_R}m_D^T\sim
\left(\begin{array}{ccc}
a^2&0&0\\
0&a^2+b^2&2ab\\
0&2ab&a^2+b^2
\end{array}\right).
\end{equation}
Rewriting the effective light neutrino mass matrix in the basis where
charged leptons are diagonal one finds
$$
\mnu\equiv
\left(\begin{array}{ccc}
a^2 +\frac{4ab}{3}+\frac{2b^2}{3}& -\frac{1}{3}b(2a+b)&-\frac{1}{3}b(2a+b)\\
-\frac{1}{3}b(2a+b)&  \frac{1}{3}b(4a-b) & a^2 -\frac{2ab}{3}+\frac{2b^2}{3} \\
-\frac{1}{3}b(2a+b)  & a^2 -\frac{2ab}{3}+\frac{2b^2}{3} & \frac{1}{3}b(4a-b)  
\end{array}\right).
$$
This matrix is fully determined by two complex parameters $a$ and $b$,
which imply three physical real parameters, namely two moduli and a
relative phase, which is the only source of leptonic CP violation in
the scheme.

We note that $\mnu$ is $\mu\leftrightarrow\tau$ invariant, so it gives
$\theta_{13}=0$ and $\sin^2\theta_{23}=1/2$ as predictions. The state
$(1,1,1)^t$ is an eigenstate of $\mnu$ with eigenvalue $a^2$, so the
neutrino mass matrix $\mnu$ is diagonalized by the tri-bimaximal
mixing matrix, leading then to $\tan^2\theta_{12}=0.5$.  The
three neutrino mass eigenvalues are
$$\{m_1, m_2, m_3\} = \{(a+b)^2,a^2,-(a-b)^2 \}.$$
Data from neutrino oscillation
experiments~\cite{fukuda:2002pe,ahmad:2002jz,araki:2004mb,Kajita:2004ga,ahn:2002up}
determine pretty well two of the three parameters on the left-hand
side~\cite{Maltoni:2004ei}, namely the solar and atmospheric
mass-square splittings. The remaining observable is precisely the
neutrino-exchange amplitude for neutrinoles double beta decay, given
by
$$\vev {m_\nu} \equiv |m_{ee}|=|a^2 +\frac{4ab}{3}+\frac{2b^2}{3}|.$$

This parameter can be given as a function of the three independent
model parameters, which we choose to express in terms of the
observables $\Dma$, $\alpha$ and the relative phase between $a$ and
$b$. The latter is directly related to the Majorana CP
phase~\cite{schechter:1980gr,Schechter:1981gk,bilenky:1980cx,Doi:1980yb}
characterizing the solar neutrino sub-system, $\phi_{12}$, in a
symmetric parametrization of the lepton mixing matrix where all phases
appear attached to the corresponding mixing
angle~\cite{schechter:1980gr,Schechter:1981gk}.

\begin{center}
\begin{figure}
\includegraphics[angle=0,height=6.5cm,width=0.48\textwidth]{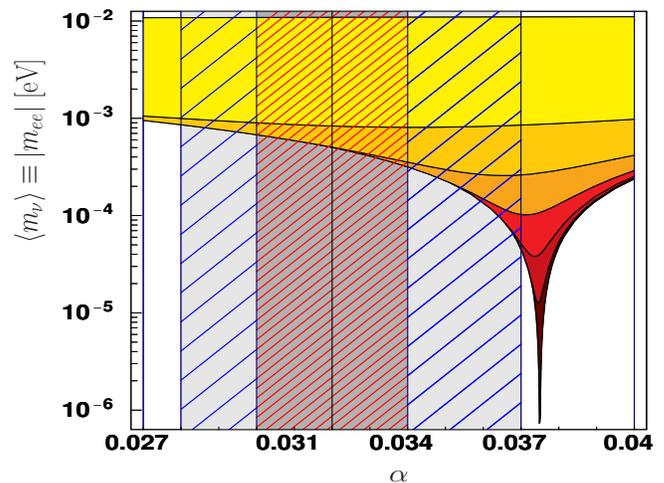}
\caption{Lower bound on the \nbb amplitude parameter $m_{ee}$ as
  function of $\alpha\equiv\Dms/\Dma$ for different values of the
  Majorana phase $\phi_{12}=-\pi/2+t$ where $t=0$~(dark brown),
  $0.001$~(brown), $0.004$~(red), $0.011$~(dark orange), $0.029$~(orange),
  $0.089$~(yellow). The 1, 2 and $3\sigma$ ranges for $\alpha$ are also shown.}
  \label{fig:mee1}
\end{figure}
\end{center}

\begin{center}
\begin{figure}
\includegraphics[angle=0,height=6.5cm,width=0.48\textwidth]{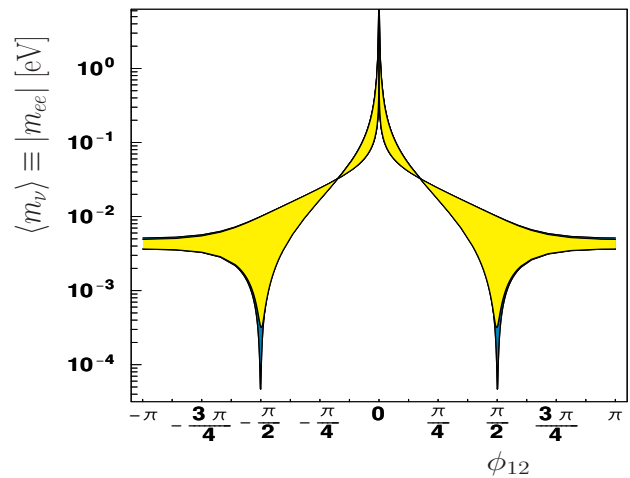}
\caption{Lower bound on the neutrinoless double beta decay amplitude
  parameter $m_{ee}$ as function of the Majorana CP phase
  $\phi_{12}$ for $\alpha$ within the $1\sigma$~(yellow) and
  $2\sigma$~(blue) ranges.}
  \label{fig:mee2}
\end{figure}
\end{center}

\begin{center}
\begin{figure}
\includegraphics[angle=0,height=6.5cm,width=0.48\textwidth]{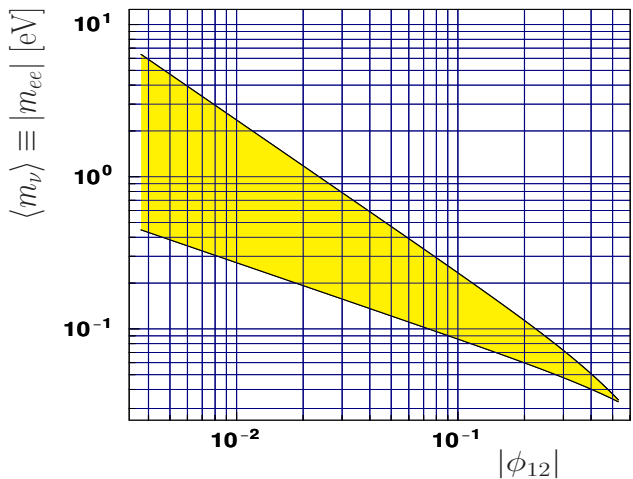}
\caption{Zoom of the region giving the maximum value
  for the lower bound on $m_{ee}$ in Fig.~\ref{fig:mee2}.}
  \label{fig:mee3}
\end{figure}
\end{center}

First we note that our scheme is compatible with negligible
neutrinoless double beta decay, vanishing at the tree level
i.~e. $m_{ee}=0$. This happens only when CP is conserved with opposite
CP parities~\cite{schechter:1981hw,Wolfenstein:1981rk} between $\nu_1$
and $\nu_2$ and for
\begin{equation}
\alpha=\frac{\Delta m_{sol}^2}{\Delta m_{atm}^2}=\frac{3}{80}=0.0375,
\end{equation}
as seen in Fig.~\ref{fig:mee1}, which is currently allowed at
3~$\sigma$.
For all other values of the CP phase the model gives a lower bound on
the neutrinoless double beta decay which we display in
Fig.~\ref{fig:mee2}, which we call the ``Niemeyer'' plot~\footnote{Due
  to its similarity to the columns of ``Palacio da Alvorada'' designed
  by this brilliant Brazilian arquitect.}.
This plot exhibits two dips characterized by very small \nbb
amplitudes, which correspond to almost full destructive interference
between opposite CP sign neutrinos $\nu_1$ and
$\nu_2$.
Notice that in the central region around the other CP conserving point
$\phi_{12}=0$ the \nbb amplitude is sizeable, and depends very
sensitively on the Majorana phase $\phi_{12}$, as displayed in
Fig.~\ref{fig:mee3}.
It is a non-trivial task to produce a consistent flavour symmetry
leading to a structure of the effective neutrino mass matrix $\mnu$
that has, at least as a first approximation, the desired
predictive pattern.

Here we suggest two possible realizations based on an $A_4$ flavour
symmetry for the neutrino mass matrix.
The discrete group $A_4$ is a relatively small and simple flavour
group consisting of the 12 even permutations among four objects. It has a
three-dimensional irreducible representation appropriate to describe
the three generations observed.  Originally, $A_4$ was proposed
\cite{Ma:2001dn,babu:2002dz} for understanding degenerate neutrino
spectrum with nearly maximal atmospheric neutrino mixing angle.  More
recently, predictions for the solar neutrino mixing angle have also
been incorporated within the so-called tri-bimaximal neutrino mixing
schemes~\cite{Altarelli:2005yx,Hirsch:2005mc,Chen:2005jm,Zee:2005ut,Ma:1,He:2006dk,Ma:2006vq}.

In our phenomenological $A_4$-based flavor symmetry schemes the
neutrino mass comes from type-I seesaw mechanism with right-handed
Majorana mass matrix proportional to the identity matrix.
In both models leptons transform as $A_4$-triplets, while the standard
Higgs is a flavour singlet~\footnote{Quarks are also flavor singlets,
  hence the quark sector is completely standard and its flavor
  structure unpredicted.}.
The lepton and scalar content of the models are specified in
Tables~\ref{tab:Multiplet1} and ~\ref{tab:Multiplet2}.
\begin{table}[h!]
\begin{center}
\begin{tabular}{|l||lll||l||lll|l|}
\hline
fields & $L_i$ & $l^c_i$ & $\nu^c_i$& $h$ & $H_i$ & $\varphi$ & $\Phi$& 
$\xi$\\
\hline
$SU(2)_L$ & 2& 1& 1& 2&2&2&2&1 \\
$A_4$  &  3& 3 &3  & 1 &3 &1&3&1\\
$Z_2$ & $+$ & $+$& $-$ & $+$ & $+$ & $-$ &$-$ &$+$ \\
\hline
\end{tabular}
\caption{Lepton multiplet structure of  model I}
\label{tab:Multiplet1}
\end{center}
\end{table}

The $A_4\times Z_2$ invariant Lagrangian characterizing the first
model is renormalizable, and given by
\begin{eqnarray*}
&&\mathcal{L}=
\lambda_0 (Ll^c)h + \lambda (Ll^c H) \\&&+ \lambda'_0 (L\nu^c)\varphi+\lambda'(L\nu^c\Phi)
+\lambda_R(\nu^c\nu^c)\xi.
\end{eqnarray*}
where the first term involves an $A_4$-invariant coupling $\lambda_0$
that provides $\alpha$ in $M^l$, while the second involves a tensor
$\lambda_{ijk}$ with components $\beta$ and $\gamma$, and similarly
for the next two terms.
Note that in this case there are additional \21 doublet Higgs scalar
bosons $H_i, \Phi_i, \varphi$ transforming non-trivially under the
flavour symmetry. We assume that these develop non-zero vacuum
expectation values (vevs), with the structure
$$
\vev{H_i} \sim (1,1,1);\quad \vev{\Phi_i} \sim (0,0,1)
$$
Similar vev alignment condition has been used in
Ref.~\cite{Altarelli:2005yp}. Note that the two zeros in $m_D$ follow
from the alignment condition $\vev{\Phi_1}=\vev{\Phi_2}=0$.

In contrast the second model contains only one \21 doublet Higgs boson
and its $A_4\times Z_4$ symmetric leading-order Lagrangian is written
as
\begin{eqnarray*}
&&\mathcal{L} = \lambda_0 (Ll^c)h\xi_1
               +\lambda(Ll^c\phi)h 
               \\&& +\lambda'_0(L\nu^c\phi')\tilde{h}
             + \lambda'(L\nu^c)h\,\xi_2
             +\lambda_R(\nu^c\nu^c)\xi_3
\end{eqnarray*}
where $\lambda_R$ is dimensionless while the others scale as inverse
mass. Note the appearance of gauge singlet scalars $\phi$, $\phi'$ and
$\xi_i$, transforming non-trivially under the flavour symmetry and
coupling non-renormalizably to the lepton doublets. Their only
renormalizable is the one giving rise to the large Majorana mass term.
We assume that these ``flavon'' fields develop non-zero vacuum
expectation values (vevs), with the structure
$$
\vev{\phi} \sim (1,1,1);\quad \vev{\phi'} \sim (0,0,1)
$$

Note that either way we obtain the desired predictive charged lepton
and neutrino mass matrices discussed above.

\begin{table}[h!]
\begin{center}
\begin{tabular}{|l||lll||l||lllll|}
\hline
fields & $L_i$ & $l^c_i$ & $\nu^c_i$& $h$ & $\phi$ & $\phi'$ & $\xi_1$& 
$\xi_2$&$\xi_3$\\
\hline
$SU(2)_L$ & 2& 1& 1& 2&1&1&1&1 &1\\
$A_4$  &  3& 3 &3  & 1 &3 &3&1&1 &1\\
$Z_4$ & 1 & $\omega^3$ & $\omega$ & 1 & $\omega$ & $\omega^3$ &$\omega$ 
&$\omega^3$ &$\omega^2$ \\
\hline
\end{tabular}
\caption{Lepton multiplet structure of model II}
\label{tab:Multiplet2}
\end{center}
\end{table}

In summary, here we have proposed two $A_4$-based flavour symmetries
leading to tri-bimaximal lepton mixing, namely
$\tan^2\theta_{\Atm}=1$, $\sin^2\theta_{\textrm{Chooz}}=0$ and
$\tan^2\theta_{\Sol}=0.5$. Although this implies a boring scenario for
upcoming long baseline oscillation
experiments~\cite{Bandyopadhyay:2007kx,Nunokawa:2007qh} aiming to
probe $\theta_{13}$ and leptonic CP violation in oscillations, we have
analysed its implications for neutrinoless double beta decay.
We have seen how the \nbb amplitude parameter $m_{ee}$ has a lower
bound which correlates with the ratio $\alpha \equiv \Dms/\Dma$ well
determined by current neutrino oscillation data, as well as with the
Majorana CP violating phase $\phi_{12}$.  Accelerator neutrino
oscillation experiments like MINOS, T2K and NOvA are expected to
improve the determination of $\alpha$ in the not-too-distant future.

For the special value $\phi_{12}=\frac{\pi}{2}$ (opposite
CP-sign neutrinos) one finds that \nbb vanishes at tree level when
$\Dms/\Dma = 3/80$. However this is only allowed at $3\sigma$, as seen
from Fig.~\ref{fig:mee1}, at $1~\sigma$ we currently have a lower
bound $|m_{ee}| \gsim$ few~$\times 10^{-4}$~eV.  For all other cases
one has a nonzero \nbb decay rate, with CP conservation with same
CP-sign neutrinos already excluded.
We have also presented in Fig.~\ref{fig:mee3} the lower bound in the
region close to $\phi_{12}=0$, corresponding to the case of same
CP-parity neutrinos, where neutrinoless double beta decay could soon
be observed.

All our considerations refer to an effective low-energy model which
assumes the vev alignment conditions, and the symmetry of the neutrino
Dirac mass matrix, a relation which holds in the framework of $SO(10)$
unification~\cite{Morisi:2007ft,Grimus:2008tm}.
A more complete picture formulated at the unified level is outside the
scope of this letter.  In principle the structure presented here can
be lifted to the $SO(10)$ level, though fitting the flavour
structure of quarks will require additional fields and/or
symmetries. 
In such more complete scenario exact tri-bimaximality would be just a
first approximation, corrections leading to calculable deviations from
the predictions reported here.  These issues will be taken up
elsewhere.

\vskip .2cm {\bf Acknowledgments} \vskip .2cm Work
supported by MEC grant FPA2005-01269, by European Commission Contracts
MRTN-CT-2004-503369 and ILIAS/N6 RII3-CT-2004-506222.

 \def\baselinestretch{1}

\end{document}